\documentclass{emulateapj}
\usepackage{xcolor}
\usepackage{bm}
\usepackage[T1]{fontenc}
\usepackage{multirow, array}

\begin{document}

\def\be{\begin{eqnarray}}
\def\ee{\end{eqnarray}}
\def\J{{\bf J}}

\title{on the power to constrain the accretion history of massive 
black holes via spin measurements by upcoming X-ray telescopes}
\author{Xiaoxia Zhang$^{1,\dagger}$, Youjun Lu$^{2,3}$, Dandan Wang,$^{4}$, and Taotao Fang$^1$}
\affil{
 ~$^1$~Department of Astronomy, Xiamen University, Xiamen, Fujian 361005, China;~$^{\dagger}$zhangxx@xmu.edu.cn; fangt@xmu.edu.cn \\
 ~$^2$~National Astronomical Observatories, Chinese Academy of
Sciences, Beijing 100012, China; luyj@nao.cas.cn \\
~$^3$~School of Astronomy and Space Sciences, University of Chinese
Academy of Sciences, No. 19A Yuquan Road, Beijing 100049, China \\
~$^4$~Department of Astronomy, School of Physics and Astronomy, Shanghai Jiao Tong University, 800 Dongchuan Road, Shanghai 200240, China
}

\begin{abstract}

The spin distribution of massive black holes (MBHs) contains rich information 
on their assembly history. However, only limited information can be extracted 
from currently available spin measurements of MBHs owing to the small sample 
size and large measurement uncertainties. Upcoming X-ray telescopes with 
improved spectral resolution and larger effective area are expected to provide 
new insights into the growth history of MBHs. Here we investigate, at a proof 
of concept level, how stringent constraints can be placed on the accretion history 
of MBHs by the spin measurements from future X-ray missions. We assume a toy 
model consisting of a two-phase accretion history composed of an initial coherent 
phase with a constant disk orientation, followed by a chaotic phase with random 
disk orientations in each accretion episode. By utilizing mock spin data generated 
from such models and performing Bayesian Markov Chain Monte Carlo simulations, 
we find that most accretion models of MBHs can be reconstructed provided that 
$\gtrsim100$ MBH spins are measured with an accuracy of $\lesssim0.1$. We also 
quantify the precision of the reconstructed parameters by adopting various combinations 
of sample sizes and spin accuracies, and find that the sample size is more crucial to 
model reconstruction once the spin accuracy reaches $\sim 0.1$. To some extent, 
a better spin accuracy will compensate for a small sample size and vice versa. Future 
X-ray missions such as the Advanced Telescope for High Energy Astrophysics and the 
enhanced X-ray Timing and Polarimetry mission, may provide spin measurements of 
$\gtrsim100$ MBHs with an uncertainty of $\sim0.04-0.1$ and will thus put strong 
constraints on the MBH growth history.
\end{abstract}
\section {introduction}
\label{sec:Intro}

Massive black holes (MBHs) in galactic centers are expected to be rotating, 
as the acquisition of angular momenta accompanies the growth of mass 
either through MBH mergers or gas accretion. It has been revealed that 
MBHs gain their masses mainly through gas accretion during the quasi-stellar 
object/active galactic nucleus (AGN) phases \citep[e.g.,][]{Yu02, Marconi04, 
Yu04, Shankar09} by comparing the local MBH mass density with the accreted 
mass density over the cosmic time via the \citet{Soltan82} argument. Accretion 
modes may leave imprints on the spin evolution of MBHs \citep[e.g.,][]{King08, 
Dotti13, Volonteri13}, and observationally determined spin distribution will in turn 
provide insights into the accretion history of MBHs. Coherent accretion with 
almost fixed disk orientation will efficiently increase the MBH spin, to an extreme 
value when the mass is $\sim2.4$ times larger than the initial mass \citep[e.g.,][]{Thorne74}. 
Instead, chaotic accretion composes a number of small episodes and in each 
episode the gas cloud infalls with a random direction of orbital angular momentum, 
and spins down the MBH because the radius of the innermost stable circular orbit 
(ISCO) is larger for a retrograde disk, thus the injection of negative angular momenta 
is more effective \citep[e.g.,][]{Bardeen72, Moderski98}.

Observationally, the spins of MBHs are mostly measured by modeling the reflected 
X-ray spectra of AGNs with a basic assumption that the accretion disk is geometrically 
thin and optically thick \citep[see][for reviews of MBH spin measurements]{Brenneman13, 
Reynolds14, Reynolds19}. In this case, the ISCO radius, which is crucial to the degree 
of relativistic effects on the X-ray spectrum in terms of broadened and skewed line profiles 
\citep[e.g.,][]{Fabian89, Laor91}, is solely determined by the MBH spin \citep{Bardeen72}, 
and the modeling of the profiles will in turn provide a measure of the MBH spin 
\citep[e.g.,][]{Tanaka95}. Currently, MBHs with relatively robust spin measurements 
amount to no more than $\sim30$, most of which are rapidly rotating \citep[][see also 
\citealt{ZL19}]{Brenneman13, Reynolds14, Vasudevan16}.

Despite the limited size of the spin sample, current data do provide some clues to the 
growth history of MBHs \citep[e.g.,][]{Dotti13, Sesana14, Li15}. In our previous work 
\citep{ZL19}, we adopted a two-phase accretion model with an initial coherent phase 
followed by a chaotic one, and obtained constraints on the MBH accretion history by 
comparing the spin-mass distributions resulting from various models with that of the  
observed samples via the two-dimensional Kolmogorov–Smirnov (2D-KS) test. We 
found that chaotic accretion is necessary and the disk-to-BH mass ratio in each 
episode is about $1\%-2\%$. By its nature, 2D-KS tests do not allow us to constrain 
models with more than two free parameters, something that one can do within the 
Bayesian framework. However, the current sample size, in addition to the error budget, 
imposes limits on performing concrete Bayesian method.

Future X-ray telescopes such as Advanced Telescope for High ENergy Astrophysics 
\citep[Athena;][]{Barret18} and the Lynx X-ray Observatory \citep[Lynx;][]{Bandler19, 
Gaskin19} will make significant improvements to the spectral resolution (to $2-3\ {\rm 
eV}$ at $\sim 6\ {\rm keV}$), and missions like the enhanced X-ray Timing and 
Polarimetry mission \citep[eXTP;][]{Zhang19}, Spectroscopic Time-Resolving 
Observatory for Broadband Energy X-rays  \citep[STROBE-X;][]{Ray19}, and the 
High-Energy X-ray Probe \citep[HEX-P;][]{Madsen18} will provide wide effective 
areas (up to $5\ {\rm m^2}$ at $\sim 6\ {\rm keV}$) with a broadband energy coverage. 
Those will greatly enlarge the current spin sample and improve the accuracy of spin 
measurements, and will set unprecedented constraints on the assembly history of MBHs. 
In this paper, we make use of mock spin samples generated from two-phase accretion 
models, to investigate how well the growth history of MBHs can be constrained with those 
future X-ray detectors.

The paper is organized as follows. In Section~\ref{sec:model}, we describe the 
accretion model and equations governing the mass and spin evolution of MBHs. 
The mock samples generated from various models are presented in 
Section~\ref{sec:data}, followed by a description of Bayesian statistics in 
Section~\ref{sec:bay}. Reconstruction results on various models are given in 
Section~\ref{sec:res}, by adopting samples of different sizes and spin accuracies. 
We discuss our results in Section~\ref{sec:dis} and present our conclusions in 
Section~\ref{sec:con}.

\section{accretion model and evolution of MBH spins}
\label{sec:model}

The current spin sample of MBHs exhibits a trend in which lighter black holes (BHs) 
mostly rotate faster and heavier ones rotate more slowly, indicating that chaotic 
accretion may play a critical role in shaping the evolution of MBHs. We therefore 
follow \citet{ZL19} and consider accretion history models composed of both coherent 
and chaotic accretion, which may be taken as an approximation to the true accretion 
history (e.g., Model A with three parameters described below). Note that the real 
accretion history of the MBHs may depend on detailed properties of their host 
galaxies and could be different among individual MBHs. For simplicity, here we only 
consider the dependence on the mass of the host galaxy and thus the MBH mass, 
since the MBH mass correlates well with the stellar mass of the host galaxy 
\citep[e.g.,][]{HR04, Hopkins07}. Our model could be approximated as the mean of 
the assembly history of those MBHs with similar host properties (stellar mass, velocity 
dispersion, etc.)

Initially, the host galaxy is probably rich in gas and the fueling of the central MBH is 
continuous and coherent, at a sub- or super-Eddington rate. With the consumption 
of gas, the MBH may later enter into a sub-Eddington chaotic accretion phase, 
composed of many small accretion episodes with random disk orientations. The 
mass of the gas cloud accreted in each chaotic episode is assumed to follow a 
power-law dependence on the MBH mass $M_\bullet$, i.e.,
\be
M_{\rm cl} (M_\bullet)=b M_\bullet   \left( \frac{M_\bullet}{10^8 M_\odot} \right)^\gamma,
\ee
where $b$ and $\gamma$ are constant parameters.  If $\gamma=0$, then the mass 
of the gas cloud scales linearly with the MBH mass; if $\gamma=-1$, then the cloud 
mass is a constant for each episode in the chaotic accretion phase. The whole cloud 
is assumed to form an accretion disk with negligible mass loss. For super-Eddington 
accretion, the accretion disk is thick in geometry, while for a sub-Eddington rate, it 
forms a standard thin disk instead \citep{Shakura73, Novikov73}.

The division of the two phases is characterized by parameter $f_{\rm c}$ such that the 
second phase starts once the MBH mass reaches a factor $f_{\rm c}$ of the final mass 
$M_{\bullet, {\rm f}}$. If $f_{\rm c}=1$, then the MBH only experiences coherent accretion; 
if $f_{\rm c} \le M_{\bullet, i}/M_{\bullet, {\rm f}}$ with $M_{\bullet, i}$ the initial MBH mass, 
then only chaotic accretion happens. $f_{\rm c}$ could be a constant, but it could also be 
dependent on the MBH final mass  \citep[e.g., see][]{Dotti13, Sesana14, ZLL19}. Here the 
final mass means the mass of a quiescent MBH at the present time ($z=0$). The dependence 
of $f_{\rm c}$ is implied by MBHs with different final masses having different accretion histories.
Below we consider three models.

Model A: $f_{\rm c}$ is a constant ($\leq 1$). $b$ and $\gamma$ are also constant 
parameters (same for the following two models).

Model B: $f_{\rm c}$ follows a power-law dependence on the MBH mass, i.e., 
\be
f_{\rm c}(M_\bullet)=f_0  \left( \frac{M_\bullet}{10^8 M_\odot} \right)^\alpha,
\label{eq-fc1}
\ee
where $f_0$ and $\alpha$ are both constant parameters and $M_{\bullet, i}/M_{\bullet, {\rm f}} 
\le f_{\rm c} \le1$. If $\alpha=0$, then it reduces to Model A.

Model C:  $f_{\rm c}$ also depends on the MBH mass but it follows a hyperbolic-tangent 
law such that $f_{\rm c}=1$ for MBHs with $M_\bullet>M_1$ and $f_{\rm c}=0$ for 
$M_\bullet<M_2$, where $M_1>M_2$, i.e., \citep[see][for similar expressions]{Gallo19}  
\be
f_{\rm c}(M_\bullet)=0.5+0.5 \tanh \left( 2.5^\beta \log \frac{M_\bullet}{M_0} \right), 
\label{eq-fc2}
\ee
where $\beta$ and $M_0$ are parameters. $M_0$ marks the mass at which the MBH 
has $f_{\rm c}=0.5$, and for convenience we introduce $m_0 \equiv \log(M_0/M_\odot)$; 
$\beta$ determines the slope of the transition from $f_{\rm c}=0$ at the low-mass end to 
$f_{\rm c}=1$ at the high-mass end, and the transition is faster for a larger $\beta$.

Regardless of the disk geometry, the mass growth of an MBH is completely determined 
by the accretion rate $\dot{M}$ and the specific energy $E_{\rm in}$ at the inner disk 
boundary $R_{\rm in}$ if the kinetic energy loss is negligible, i.e., 
\be
\frac{d M_\bullet}{d t}=f_{\rm Edd} \frac{E_{\rm in}}{1-E_{\rm in}} 
\frac{M_\bullet}{t_{\rm Edd}}, 
\label{eq:dmdt}
\ee
where $f_{\rm Edd}$ is the Eddington ratio, and $t_{\rm Edd}=4.5\times10^8\ {\rm yr}$ 
is the Eddington timescale. The inner boundary of the disk is different for thick and thin 
disks. For thin disks, $R_{\rm in}$ is the ISCO radius, which is a monotonically determined 
by the MBH spin, i.e., $R_{\rm in} = 3+Z_2 \mp[(3-Z_1)(3+Z_1+2Z_2)]^{1/2}$, where $Z_1$ 
and $Z_2$ are functions of spin, and the upper and lower cases of `$\mp$' are respectively 
for prograde and retrograde orbits \citep{Bardeen72}. For thick disks, $R_{\rm in}$ is between 
the marginally bound and marginally stable orbits \citep{koz78}. Below we only consider 
accretion rate capped by the Eddington limit and thus a thin-disk in geometry. As the accretion 
rate mainly affects the accretion time instead of the spin-mass evolutionary tracks, this will 
have little effect on our main results. With the assumption that the Eddington ratio is 
logarithmically dependent on the accretion rate, $R_{\rm in}$ can be derived through 
interpolation \citep[for details see][]{ZL19}. At a given spin, the specific energy $E_{\rm in}$ 
(and also the specific angular momentum $\Phi_{\rm in}$ in Equation~(\ref{eq:djdt})) at any 
$R_{\rm in}$ can be obtained \citep{Bardeen72}. The mass-to-energy conversion efficiency 
$\eta$ is related to $E_{\rm in}$ as $\eta=1-E_{\rm in}$  if the kinetic energy loss is neglected.

For the general case of an inclined accretion disk with respect to the MBH spin, owing to the 
frame-dragging effect, the inner disk will be bent to the equatorial plane of the MBH and the 
outer disk maintains the original orientation \citep{Bardeen75}. Then the evolution of MBH 
angular momentum vector $\J_\bullet$ is governed by
\be
\frac{d \J_\bullet}{d t} = \dot{M} \frac{G M_\bullet}{c}\Phi_{\rm in}\hat{\bf
l}+\frac{4\pi G}{c^2}\int_{\rm disk} \frac{{\bf L}\times \J_\bullet}{R^2} d R, 
\label{eq:djdt}
\ee 
where $\Phi_{\rm in}$ is the specific angular momentum at $R_{\rm in}$, $\hat{\bf l}$ is a unit 
vector paralleled with $\J_\bullet$, and ${\bf L}$ is the angular momentum vector of the disk 
per unit area. The first term on the right side denotes the angular momentum change due to 
plunge-in of material at $R_{\rm in}$, and only leads to the modification of the spin module, 
while the second term describes the gravito-magnetic interaction between the disk and MBH, 
and only causes spin direction change. Equations~(\ref{eq:dmdt}) and (\ref{eq:djdt}) are solved 
via the adiabatic approximation, i.e., the disk transits through a sequence of steady warped 
states over time interval $\delta t \ll t_{\rm al}$, where $t_{\rm al}$ is the alignment timescale 
\citep{Perego09}. If the disk and MBH angular momenta already align with each other, then 
the spin direction does not change for the rest of the episode, and Equation~(\ref{eq:djdt}) 
reduces to 
\be
\frac{da}{dt}  = ( \Phi_{\rm in}- 2a E_{\rm in} ) \frac{f_{\rm Edd}}{(1-E_{\rm in}) \ t_{\rm Edd}},
\label{eq:dadt}
\ee 
where $a$ is the dimensionless spin parameter defined as $|a|=c|{\bf J}_\bullet|/(GM^2_\bullet)$, 
and $a$ is positive (negative) if the disk is corotating (counterrotating) around the MBH.

In our calculation, the initial mass of MBHs is fixed at $10^5 M_\odot$, the final mass is 
randomly selected over $10^6 M_\odot$ and $10^{10} M_\odot$ in logarithmic space, 
and the initial spin is randomly drawn from a uniform distribution between $0$ and $1$. 
The accretion rate of the initial coherent phase is fixed at $\dot{m} \equiv \dot{M}/\dot{M}_{\rm 
Edd}=0.3$ where $\dot{M}_{\rm Edd}=16L_{\rm Edd}/c^2$ with $L_{\rm Edd}$ the Eddington 
luminosity, and the Eddington ratio $f_{\rm Edd}$ in each chaotic phase is set to be a constant 
that is randomly drawn from a Gaussian distribution with a mean of $0.1$ and standard deviation
of $0.3$ in logarithmic space. The maximum spin is assumed to be the canonical value of $0.998$, 
i.e., $|a| \le 0.998$ \citep{Thorne74}.

\begin{figure}
\centering
\includegraphics[width=3.4in]{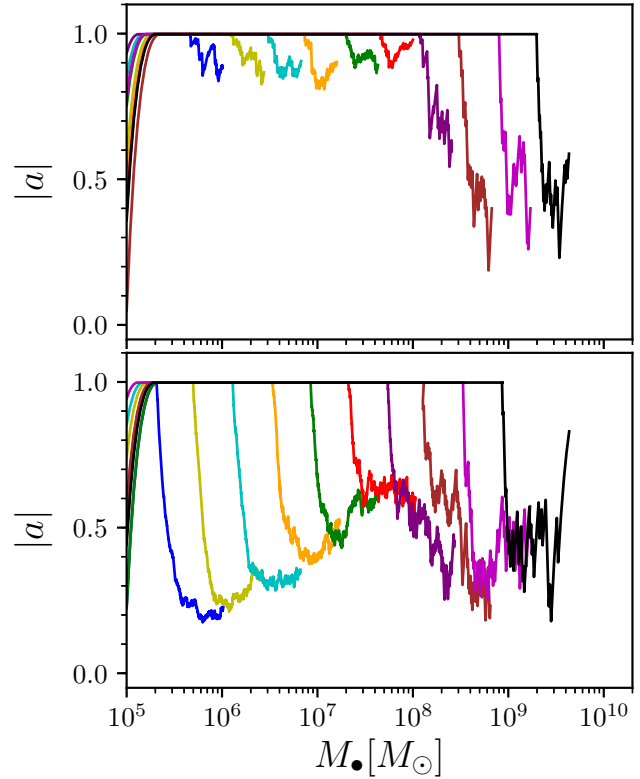}
\caption{Examples of spin evolutionary curves for MBHs undergoing two-phase accretion 
with constant $f_{\rm c}$ and linearly scaled $M_{\rm cl}$ with the MBH mass, i.e., $(b, 
\gamma, f_{\rm c})=(0.012, 0.00, 0.46)$ for the upper panel, and $(b, \gamma, f_{\rm c})
=(0.005, 0.50, 0.20)$ for the lower panel. Each of the MBHs has a fixed initial mass of 
$10^5 M_\odot$ and a random initial spin over $0$ and $1$. Different colors represent 
MBHs with different final masses.}
\label{fig-cv1}
\end{figure}

\begin{figure}
\centering
\includegraphics[width=3.4in]{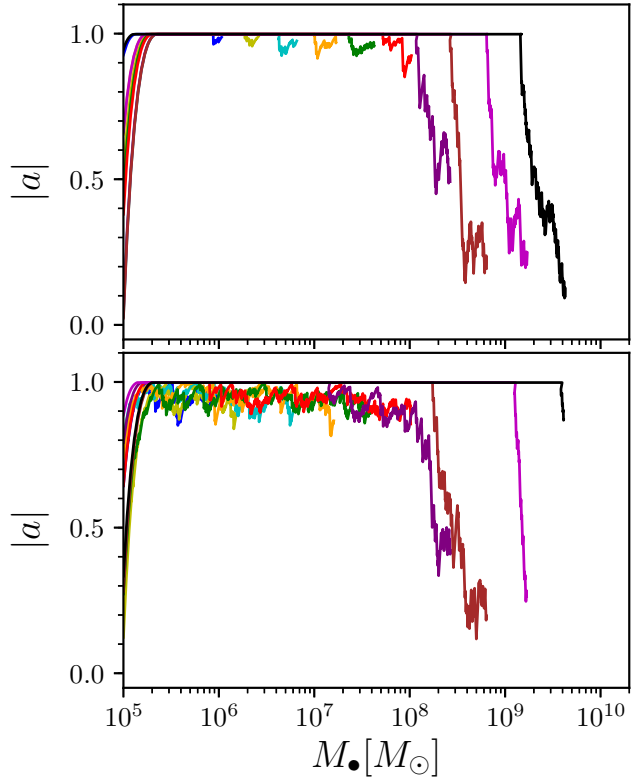}
\caption{Examples of spin evolution for MBHs experiencing two-phase accretion model with 
power-law dependence of $f_{\rm c}$ on the MBH mass, i.e., $(b, \gamma, f_0, \alpha)=(0.010, 
-0.20, 0.50, -0.10)$ for Model B (upper panel), and with hyperbolic-tangent form of $f_{\rm c}$ 
with $(b, \gamma, m_0, \beta)=(0.010, -0.20, 9.00, 1.00)$ for Model C (lower panel). The legend 
is the same as that in Figure~\ref{fig-cv1}.}
\label{fig-cv2}
\end{figure}

We first consider the best-fit model implied by our previous work, i.e., Model A with $(b, \gamma, 
f_{\rm c})=(0.012, 0.00, 0.46)$, and examples of spin evolution of MBHs are shown in the top 
panel of Figure~\ref{fig-cv1}. As demonstrated in \citet{ZL19}, the initial sharp increase of the 
spin before the mass doubles is due to quick alignment of MBH spin to the disk angular momentum. 
Then the spin maintains the maximum value of $0.998$ until the second chaotic phase decreases 
it somehow. More massive MBHs can be spun down to lower values because disk-to-BH angular 
momenta ratio decreases with increasing MBH mass as $J_{\rm d}/J_{\bullet} \propto M^{-12/25}_{\bullet}$. 
For arbitrary initial configurations, the criterion for final anti-alignment, i.e., $\cos\beta<-J_{\rm d}/J_{\bullet}$ 
with $\beta$ being the initial angle between the two angular momenta \citep{King05}, is more 
frequently satisfied for high-mass BHs, and the probability is $\sim0.5$ if $J_{\rm d}/J_{\bullet} \ll 1$.

We also consider alternative values of the parameters in Model A, i.e., $(b, \gamma, f_{\rm c})
=(0.005, 0.50, 0.20)$, which means that the second chaotic phase is more significant to the 
MBH growth, and that $J_{\rm d}/J_{\bullet}$ in each chaotic episode is smaller for low-mass 
BHs and larger for high-mass ones, compared to the former case. The spin evolution is shown 
in the lower panel of Figure~\ref{fig-cv1}, and the evolution trend is easy to understand. A smaller 
cloud and thus a smaller $J_{\rm d}/J_{\bullet}$ leads to a more rapid and efficient decline in the 
MBH spin, and a combination with a smaller $f_{\rm c}$ leads to the low spins for low-mass BHs. 
For high-mass BHs, their lower spins are results of competition between a larger cloud and a more 
significant chaotic phase. This results in a feature in the spin-mass distribution in which low-mass 
and high-mass BHs have low spins while intermediate-mass BHs have relatively high spins (see 
also the lower panel of Figure~\ref{fig-data3}).

We then consider Model B and set $(b, \gamma, f_0, \alpha)=(0.010, -0.20, 0.50, -0.10)$. 
This results in the spin evolution of MBHs presented in the upper panel of Figure~\ref{fig-cv2}. 
The spin decrease in the chaotic phase for MBHs with final masses $<10^8 M_\odot$ is not 
as significant as that in the top panel of Figure~\ref{fig-cv1} because of a larger $M_{\rm cl}$ 
and thus $J_{\rm d}/J_{\bullet}$ in each episode, while the spin can decrease to lower values 
for $>10^8 M_\odot$ MBHs owing to a smaller $M_{\rm cl}$. The lower panel of Figure~\ref{fig-cv2} 
is for Model C with $f_{\rm c}$ with $(b, \gamma, m_0, \beta)=(0.010, -0.20, 9.00, 1.00)$. Now 
low-mass BHs ($<10^8 M_\odot$) are not that extremely rotating, with $|a|\gtrsim0.8$ for most 
of their lifetimes. This occurs because low-mass BHs experience pure chaotic accretion with 
$f_{\rm c}\sim0$. The most massive BHs undergo a brief chaotic phase, and the spin remains 
at a high value (see the rightmost black curve).

\section{mock data}
\label{sec:data}

Precise measurement of MBH spin via X-ray reflection spectroscopy relies on both the spectral 
resolution around $6\ {\rm keV}$ to resolve the profile of the Fe K$\alpha$ line and high sensitivity 
in the $10-100\ {\rm keV}$ band to characterize the reflection spectrum \citep[e.g.,][]{Garcia19}. 
Future missions are expected to make improvements in both the sample size and accuracy of 
spin measurements \citep{Miller07}. For instance, the X-ray Integral Field Unit (X-IFU) instrument 
on board Athena could reach a mean spin measurement error of $\sim0.04$ \citep{Barret19}.
With a similar spectral energy resolution, Lynx may also achieve similar precisions. The eXTP 
incorporates the large area detector and the spectroscopic focusing array, and will make it 
possible to carry out detailed broad Fe line modeling on a sample of $\sim400$ AGNs at different 
redshifts to measure MBH spins with an accuracy of $10\%-20\%$ \citep[e.g., ][]{Rosa19}. The 
concept mission STROBE-X has an effective area of $\sim5\ {\rm m^2}$ at the iron line, which will 
lead to an accuracy of $<10\%$ for spin measurement \citep{Ray19}. In a word, at least hundreds 
of MBH spins are expected to be measured with an accuracy of $\lesssim0.1$ in the future. We 
therefore consider samples including $100-1000$ MBHs with spin errors of $0.04-0.1$. The mock 
spin samples are generated from the accretion model described in Section~\ref{sec:model}.

We follow the mass and spin evolutionary tracks for a population of $1000$ MBHs according 
to the settings described in Section~\ref{sec:model}. Those spin-mass evolutionary curves are 
treated as templates, from which we randomly select a number of mock MBHs with known masses 
and spins as the mock data. We additionally assign a measurement error of $0.3$ dex to the MBH 
mass according to the uncertainty of virial mass estimates \citep[e.g.,][]{Shen08, Shen11}. The 
accuracy of spin measurement could be different among various future X-ray detectors, and may 
also depend on the intrinsic spin value. Here we consider two cases of spin measurement errors, 
i.e., $0.1$ and $0.04$.

\begin{figure}
\centering
\includegraphics[width=3.4in]{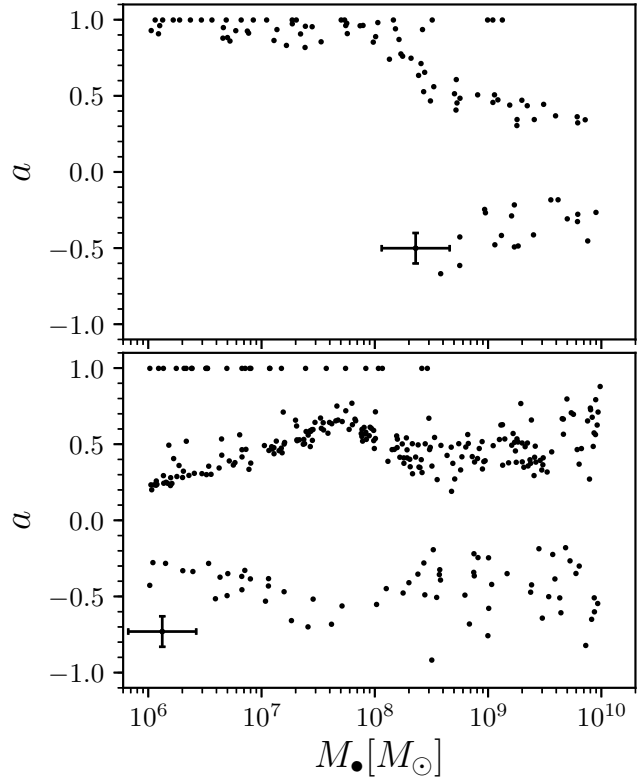}
\caption{Spin-mass distribution of the mock sample generated from Model A with $(b, \gamma, 
f_{\rm c})=(0.012, 0.00, 0.46)$ (upper panel) and $(b, \gamma, f_{\rm c})=(0.005, 0.50, 0.20)$ 
(lower panel), including $100$ and $300$ objects, respectively.  The error bar in each panel is 
to indicate the assumed measurement error for the whole sample, i.e., $0.1$ for spin and $0.3$ 
dex for mass. }
\label{fig-data3}
\end{figure}

\begin{figure}
\centering
\includegraphics[width=3.4in]{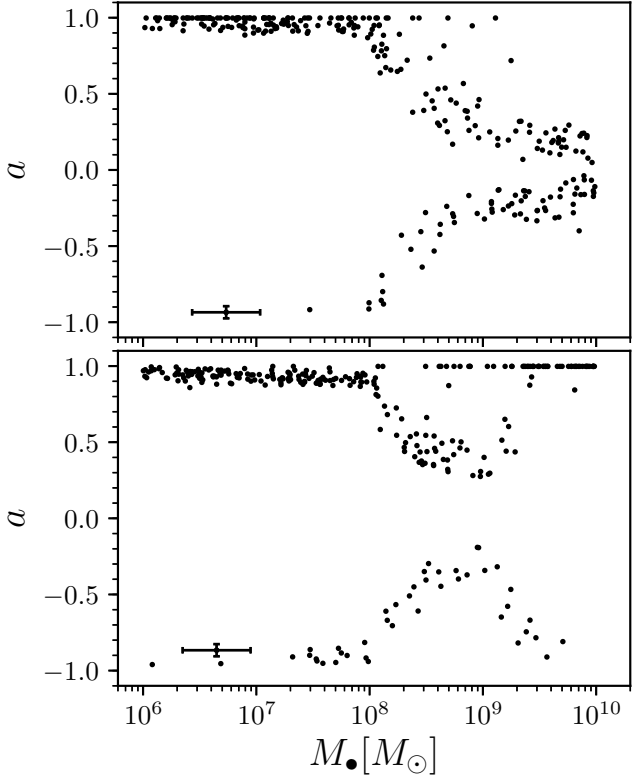}
\caption{Spin-mass distribution of the sample generated from Model B with $(b, \gamma, f_0, 
\alpha)=(0.010, -0.20, 0.50, -0.10)$ (upper panel), and from Model C with $(b, \gamma, m_0, 
\beta)=(0.010, -0.20, 9.00, 1.00)$. The two sets of samples both include $300$ objects with an 
assumed spin measurement error of $0.04$.}
\label{fig-data4}
\end{figure}

The upper panel of Figure~\ref{fig-data3} shows the spin-mass distribution of $100$ mock objects 
generated from Model A with $(b, \gamma, f_{\rm c}) =(0.012, 0.00, 0.46)$, which exhibits high spins 
($|a|\gtrsim0.8$) for low-mass BHs and a broad range of spins ($|a| \sim 0.2-1$) for high-mass ones. 
This is a natural result of the spin evolutionary curves shown in the upper panel of Figure~\ref{fig-cv1}. 
There is also a fraction of counterrotating BHs, and the fraction is higher for heavier ones, because a 
more massive BH has a higher probability of reaching a final anti-alignment configuration owing to a 
smaller disk-to-BH momenta ratio (see also the caption of Figure~\ref{fig-cv1}).

The spin distribution of $300$ sources in the bottom panel of Figure~\ref{fig-data3} is Model A 
with $(b, \gamma, f_{\rm c})=(0.005, 0.50, 0.20)$, which mainly concentrates at intermediate 
spins ($|a|\sim0.2-0.7$). The distribution also follows the spin evolutionary tracks shown in the 
lower panel of Figure~\ref{fig-cv1}. A larger fraction of negative spins is found for low-mass BHs 
compared to those in the top panel, which is attributed to smaller $J_{\rm d}/J_{\bullet}$, thus 
there is higher probability of reaching an anti-alignment configuration.

For models with mass-dependent $f_{\rm c}$, the upper panel of Figure~\ref{fig-data4} shows 
a sample of size $100$ generated from Model B with $(b, \gamma, f_0, \alpha)=(0.010, -0.20, 
0.50, -0.10)$. The lower panel is for Model C with $(b, \gamma, m_0, \beta) =(0.010, -0.20, 
9.00, 1.00)$, including $300$ sources. The spin distributions of those samples simply follow 
the spin evolutionary tracks in Figure~\ref{fig-cv2}, with a fraction of negative spins depending 
on disk-to-BH angular momenta ratios and their initial configurations.

\section{Bayesian reconstruction of MBH accretion history}
\label{sec:bay}

With the mock sample of MBH spin measurement generated in Section~\ref{sec:data}, 
the parameters associated with the accretion model are reconstructed by performing 
the Bayesian Markov Chain Monte Carlo (MCMC) simulation. The model parameters 
are denoted as ${\bm\theta}$, and $D$ represents the mock data we generated in 
Section~\ref{sec:data}.

The observations are assumed to be uncorrelated, therefore the probability of measuring
$n_i$ objects in a particular $i$th bin $\Delta M_\bullet \Delta a$ follows a Poisson distribution, 
i.e., \citep[see also][]{Sesana11, Sesana14},
\be
p(n_i)=\frac{\lambda^{n_i}_i e^{-\lambda_i}}{n_i!},
\ee
where $\lambda_i$ is the events number in the $i$th bin expected by the model. If we divide 
the parameter space (spin vs. mass) into $K$ bins, then samples in the $i$th bin can be easily 
counted. To further consider measurement errors of MBH mass and spin, we assume Gaussian 
distributions for $\log M_\bullet$ and $a$ around the mean values, with standard deviations of 
the assumed measurement errors. The probability distribution of spin is cut off at $a=\pm 0.998$ 
and normalized to $1$. The overall likelihood $p(D|{\bm\theta})$ of seeing these data under 
parameters ${\bm\theta}$ is the product of the Poisson probabilities, i.e.,
\be
p(D|{\bm\theta})=\prod^K_{i=1} \frac{(\lambda_i)^{n_i} e^{-\lambda_i}}{n_i!}.
\label{eq-like}
\ee
The logarithmic form is more frequently used, i.e.,
\be
\ln p(D|{\bm\theta})=\sum^K_{i=1} n_i \ln (\lambda_i)-\lambda_i-\ln (n_i!).
\ee
With the consideration of measurement errors, $n_i$ may not be an integer, and the factorial 
cannot be computed. However, we note that the last term $-\ln (n_i!)$ is independent of models, 
and can therefore be neglected. According to Bayesian inference, the posterior distribution of 
${\bm\theta}$ can be inferred through the likelihood and prior, i.e., 
\be
p({\bm\theta}|D) \propto p(D|{\bm\theta}) p({\bm\theta}).
\ee

\section{results}
\label{sec:res}

\subsection{Reconstruction of the Three-parameter Model}
\label{sec:par3}

As for the reconstruction of accretion models, we first adopt the mock data produced from 
Model A with $(b, \gamma, f_{\rm c})=(0.012, 0.00, 0.46)$ presented in the upper panel of 
Figure~\ref{fig-data3}. The model parameters ${\bm\theta}=(b, \gamma, f_{\rm c})$ are then 
set free, and we solve the equations governing the mass and spin evolution and randomly 
select $300, 000$ mock objects. During this process, the accretion rate $\dot{m}$ in the 
coherent phase is randomly chosen between $0.1$ and $1$ to make the thin-disk accretion 
model valid. This choice is to make sure that spin measurements for MBHs are possible for 
this phase. The Eddington ratio in the chaotic phase is assumed to be moderate, i.e., $0.01\leq 
f_{\rm Edd} \leq 1$ and the distribution is uniform in the logarithmic space over this range. 
For each set of parameters, we select $300, 000$ mock objects and compare the spin-mass 
distribution of those objects and the mock sample presented in Section~\ref{sec:data} via the 
Bayesian technique.

Note here that the accretion history model we adopted to fit the data is the same model that 
generates the mock spin sample, in order to demonstrate that the accretion history can be 
well reconstructed. In reality, however, only one among a number of possible models that 
can best match the observational data is the best approximation to the true but unknown 
prior accretion history of MBHs. In Section~\ref{sec:dis1} we will further investigate that 
whether a mock spin sample generated by one accretion history model can also be fit by 
a different one in order to show whether different accretion history models can be well 
distinguished from each other via the resulting spin distributions.

\begin{figure*}
\centering
\includegraphics[width=4.7in]{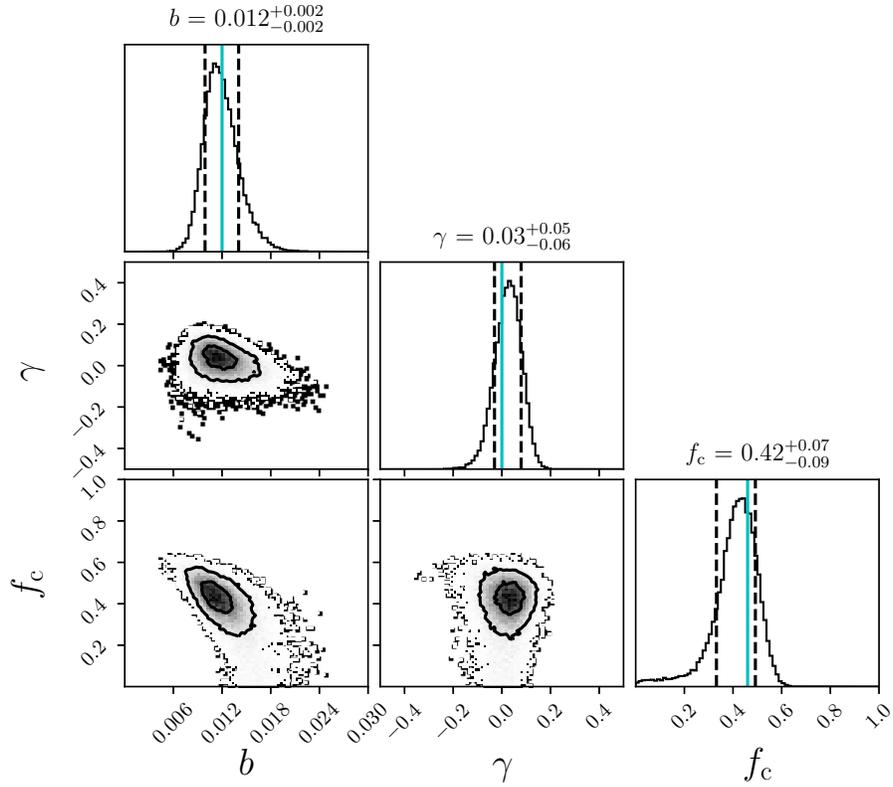}
\caption{Joint and posterior distributions of the parameters involved in Model A. The sample 
adopted (shown in the upper panel of Figure~\ref{fig-data3}) is generated from the model with 
$(b, \gamma, f_{\rm c})=(0.012, 0.00, 0.46)$ (vertical cyan lines), including $100$ sources with 
an assumed spin measurement error of $0.1$. The contours show $68\%$ and $95\%$ confidence 
levels, and each pair of dashed lines encloses a $1\sigma$ region of the parameter.}
\label{fig-case_par3}
\end{figure*} 

\begin{figure*}
\centering
\includegraphics[width=4.7in]{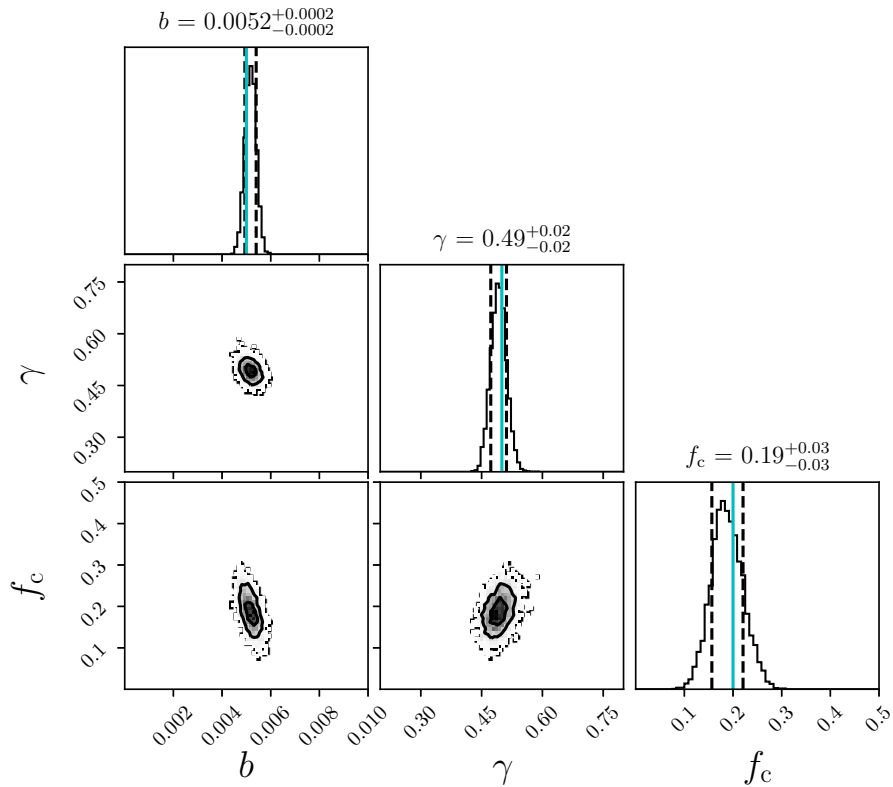}
\caption{Same as Figure~\ref{fig-case_par3} but the sample applied (shown in the lower panel of  
Figure~\ref{fig-data3}) contains $300$ MBHs, generated from Model A with $(b, \gamma, f_{\rm c})
=(0.005, 0.50, 0.20)$ (cyan lines).}
\label{fig-par3_b}
\end{figure*} 

Figure~\ref{fig-case_par3} shows constraints on the model parameters, by adopting a sample 
consisting of $100$ MBHs with a spin measurement error of $0.1$ (upper panel of Figure~\ref{fig-data3}). 
The model parameters are well reproduced by comparing the $1\sigma$ region (dashed lines) 
and the input values (cyan lines) shown in the figure. The posterior distributions of the model 
parameters are mostly Gaussian, except for $f_{\rm c}$, which has a tail on the left side. This 
occurs because pure chaotic accretion ($f_{\rm c} \sim 0$) can also result in an intermediate-to-high 
spin distribution, as demonstrated in \citet{ZL19}. A larger sample may be required to remove the tail.

We therefore apply a sample with $300$ mock objects in the the lower panel of Figure~\ref{fig-data3}, 
which also has a spin accuracy of $0.1$, generated from Model A with $(b, \gamma, f_{\rm c})=(0.005, 
0.50, 0.20)$. The model is reconstructed in Figure~\ref{fig-par3_b}. Now there is no tail for the probability 
distribution of $f_{\rm c}$, and the parameters are confined in a narrower region compared to 
Figure~\ref{fig-case_par3}. Our result implies that with a sample of size $\sim300$ and spin error of 
$\sim 0.1$, the simple three-parameter model can be perfectly reconstructed, with the $b$ parameter 
best constrained with an accuracy of $0.0002$.

\subsection{Reconstruction of Four-parameter Models}

\begin{figure*}
\centering
\includegraphics[width=4.7in]{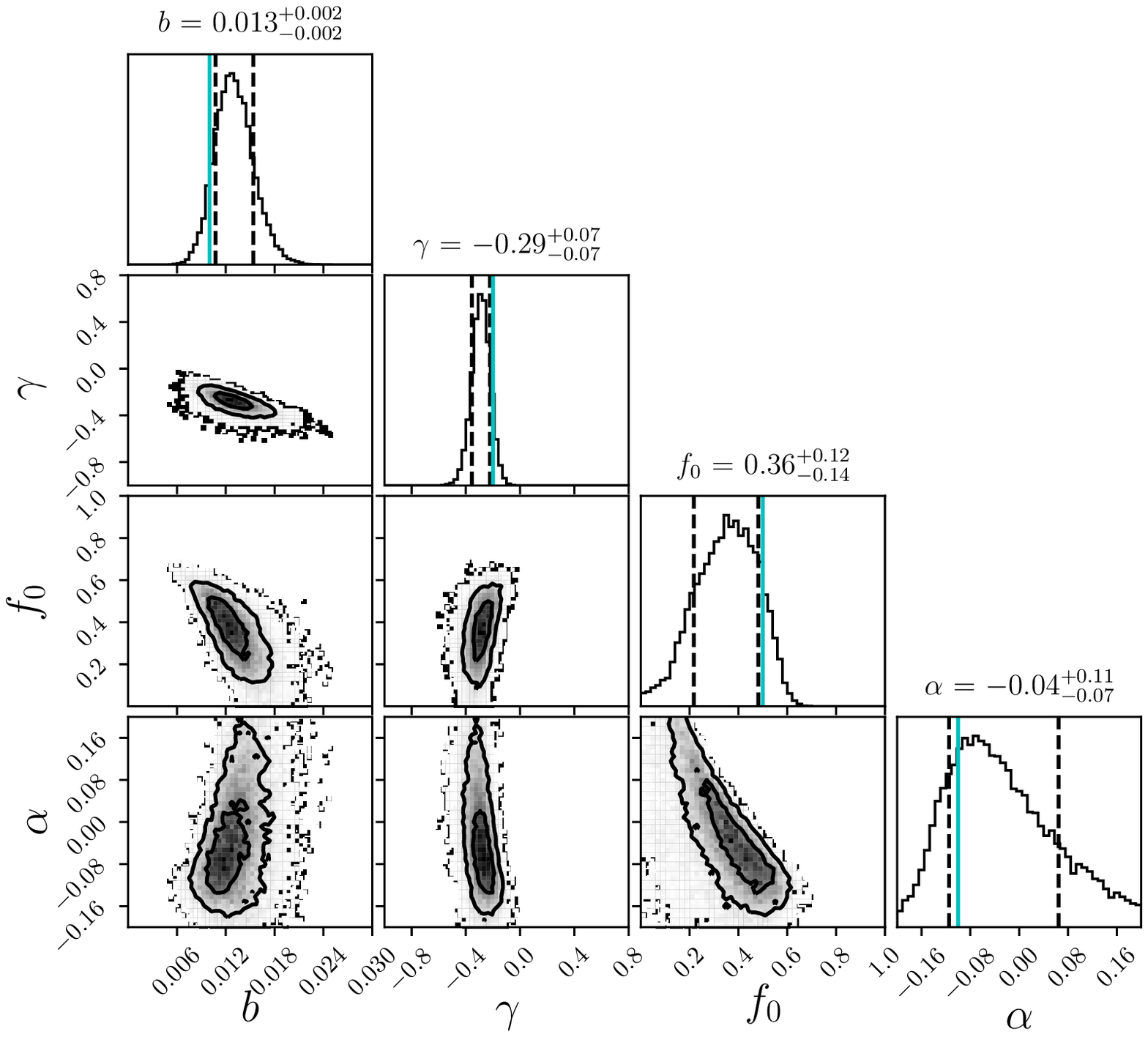}
\caption{Joint and posterior distributions of the parameters involved in the accretion Model B. 
The sample adopted includes $300$ sources with assumed spin measurement error of $0.04$ 
(upper panel of Figure~\ref{fig-data4}), generated from the model with $(b, \gamma, f_0, \alpha)
=(0.010, -0.20, 0.50, -0.10)$ (cyan lines). }
\label{fig-pl}
\end{figure*}

The results in Section~\ref{sec:par3} indicate that  Model A with constant $f_{\rm c}$ can be very 
well reproduced provided that a sample of $300$ MBH spins is measured with an error of $0.1$. 
Here we relax $f_{\rm c}$ to vary among MBHs of different masses, and consider Model B with 
power-law dependence $f_{\rm c}$. The mock sample adopted is that shown in the upper panel 
of Figure~\ref{fig-data4}, with an assumed spin measurement error of $0.04$. The Bayesian 
constraint on the model parameters ${\bm\theta}=(b, \gamma, f_0, \alpha)$ is shown in Figure~\ref{fig-pl}.

As can be seen, constraints on the model parameters are not that stringent, compared to the case 
of Model A, despite the fact that more accurate spin data are applied. The posterior distributions 
of parameters $f_0$ and $\alpha$ cover broad regions and the input values of all the parameters 
are marginally consistent with the corresponding $1\sigma$ boundary of the distribution. This might 
be because the sample is generated from the model with weak dependence of $f_{\rm c}$ on the 
MBH mass, i.e., $f_{\rm c} \propto M^{-0.1}_\bullet$, which is set to make sure that $f_{\rm c}$ does 
not exceed unity. The small difference of $f_{\rm c}$ among the different MBHs results in a bad 
constraint on $f_{\rm c}$ (in terms of $f_0$ and $\alpha$), which in turn causes a less stringent 
constraint on $M_{\rm cl}$ (in terms of $b$ and $\gamma$) because of degeneracy between 
$M_{\rm cl}$ and $f_{\rm c}$.

\begin{figure*}
\centering
\includegraphics[width=4.7in]{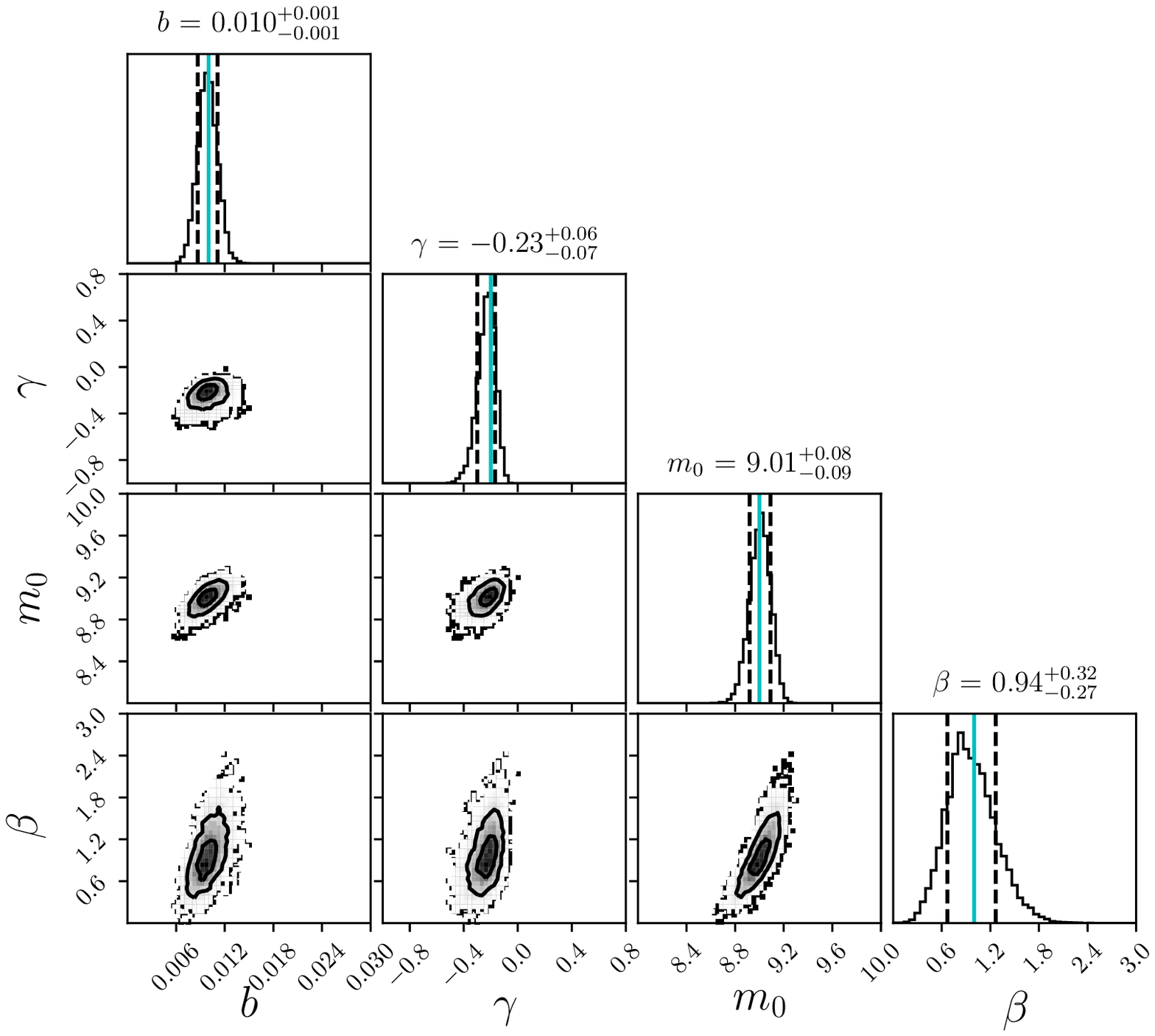}
\caption{Joint and posterior distributions of the parameters involved in the accretion Model C. 
The sample adopted includes $300$ sources with an assumed spin measurement errors of 
$0.04$ (lower panel of Figure~\ref{fig-data4}), generated from the model with $(b, \gamma, 
m_0, \beta)=(0.010, -0.20, 9.00, 1.00)$ (cyan lines). }
\label{fig-n300_004}
\end{figure*}

We therefore consider Model C with hyperbolic-tangent form of $f_{\rm c}$. By adopting the sample 
with $300$ mock objects shown in the lower panel of Figure~\ref{fig-data4}, the model parameters 
${\bm\theta}=(b, \gamma, m_0, \beta)$ are constrained, as shown in Figure~\ref{fig-n300_004}. The 
model parameters are now confined in a much narrower region as compared to the case of Model B 
shown in Figure~\ref{fig-pl}. In addition, the input values of the model parameters are well within the 
$1\sigma$ confidence regions, suggesting a successful reconstruction of the accretion model.

\subsection{The Role of Sample Size and Spin Accuracy}
\label{sec:tanh}

There is great interest in exploring the full space of the sample size and spin accuracy. Here we 
simply consider two cases of spin accuracy, i.e., $0.1$ and $0.04$, which can be approximated 
as the capacity of eXTP and Athena, respectively. For each of the two cases, we generate samples 
including $100$, $300$, and $1,000$ MBHs from Model C. Similarly, we are able to obtain the 
posterior distributions of the model parameters. 

A comparison of the reconstructed results with the input values of the model parameters is summarized 
in Table~\ref{tbl-1}, which also includes results from Models A and B. For most of the cases listed there, 
the $1-\sigma$ confidence regions of the MCMC sampling are in line with the input values of the parameters. 
This is more readily satisfied if the mock sample applied has a smaller size and lower spin accuracy because 
the posterior distribution of the model parameter covers a broader range. An unexpected case is that when 
a sample of $1000$ objects is applied to Model C, the reconstructed $\beta$ deviates more from the mean 
value despite the fact that the standard deviation is smaller toward a larger sample size and better spin 
accuracy. One reason may be that the resulting spin distribution is still not sensitive enough to parameter 
$\beta$, compared to the other three parameters, which is implied by their error bars for cases with source 
numbers of $100$ and $300$. Other factors such as degeneracy between parameters and Poisson statistics 
may also contribute to this result. Our results indicate that a larger sample and better spin accuracy in the 
future will call for improvement in the accretion model of MBHs.

\begin{figure*}
\centering
\includegraphics[width=5in]{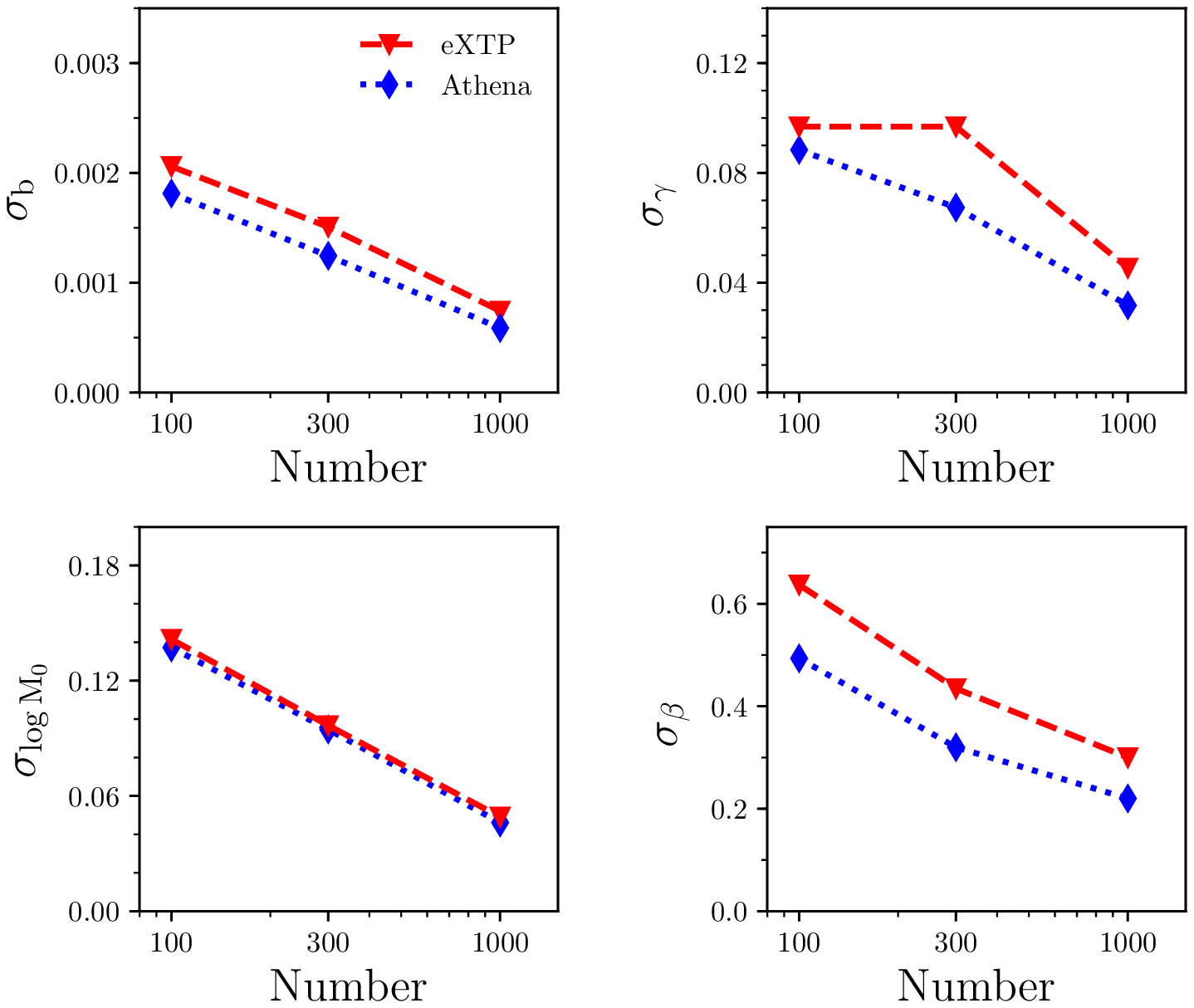}
\caption{Constraint accuracy of the parameters ($b$, $\gamma$, $M_0$, and $\beta$) involved 
in the accretion Model C, as a function of the size of the mock sample, with assumed spin 
measurement errors of $0.1$ (red triangles) and $0.04$ (blue diamonds), roughly corresponding 
to the capacity of eXTP and Athena, respectively. The symbols are connected with dashed and 
dotted lines with corresponding colors to guide the eye. The constraint accuracy is quantified by 
the standard deviation of the MCMC sampling.}
\label{fig-acc}
\end{figure*}

In addition to enabling comparison of the mean with the input value, $1-\sigma$ error is another 
quantity to measure the reconstruction of the accretion model. The standard deviations of the 
posterior distribution are treated as the constraint accuracy of the parameter, and its dependence 
on the sample size and spin precision are shown in Figure~\ref{fig-acc}. The constraint accuracies 
of the parameters are mostly improved, as expected, with increasing object numbers of the mock 
sample and with improving accuracy of spin measurements. Constraints on the parameter $M_0$ 
(lower left panel) seem to not be sensitive to the spin accuracy, indicating that with a spin accuracy 
of $0.1$, parameter $M_0$ is already well reconstructed. This is determined by its intrinsic properties 
because the resulting spin distribution is quite sensitive to $\beta$ and it converges fast in the MCMC 
sampling. In the parameter space explored here, a larger sample size will somehow compensate for a 
lower spin accuracy, and vice versa.

\begin{table*}
\begin{center} 
\renewcommand{\arraystretch}{1.8}
\caption{Comparison of Reconstructed Results and Input Values of the Parameters Associated with 
Different Versions of the Two-phase Accretion Model, by Utilizing Mock Samples of Various Sizes 
and Spin Accuracies. }
\label{tbl-1}
\begin{tabular}{ccccc}
\tableline
\tableline
Model: Parameters 	  &  Number	&  Spin Error     & Input Values    & Reconstruction Results    \\
\hline
\multirow{2}{*}{A:  $(b, \gamma, f_{\rm c})$} 	&100		& 0.1		
& $(0.012, 0.00, 0.46)$	& $(0.012^{+0.002}_{-0.002}, 0.03^{+0.05}_{-0.06}, 0.42^{+0.07}_{-0.09})$ 		\\
												& 300	& 0.1		
& $(0.0050,0.50,0.20)$	& $(0.0052^{+0.0002}_{-0.0002}, 0.49^{+0.02}_{-0.02}, 0.19^{+0.03}_{-0.03})$  		 \\
\hline
B: $(b, \gamma, f_0, \alpha)$			& 300	& 0.04	
& $(0.010, -0.20, 0.50, -0.10)$	& $(0.013^{+0.002}_{-0.002}, -0.29^{+0.07}_{-0.07}, 0.36^{+0.12}_{-0.14}, -0.04^{+0.11}_{-0.07})$ 	 \\
\hline
\multirow{6}{*}{C: $(b, \gamma, m_0, \beta)$} 		&100		& 0.1	
& \multirow{6}{*}{$(0.010, -0.20, 9.00, 1.00)$}	& $(0.010^{+0.002}_{-0.002}, -0.09^{+0.08}_{-0.11}, 9.06^{+0.13}_{-0.15}, 0.94^{+0.71}_{-0.60})$ 	 \\	
					&100		& 0.04	&  & $(0.010^{+0.002}_{-0.002}, -0.16^{+0.08}_{-0.10}, 9.06^{+0.13}_{-0.15}, 0.96^{+0.54}_{-0.43})$ 		 \\
					&300		& 0.1		& & $(0.010^{+0.002}_{-0.001}, -0.22^{+0.08}_{-0.11}, 9.04^{+0.09}_{-0.10}, 1.16^{+0.45}_{-0.39})$ 		  \\
					&300		& 0.04	& & $(0.010^{+0.001}_{-0.001}, -0.23^{+0.06}_{-0.07}, 9.01^{+0.08}_{-0.09}, 0.94^{+0.32}_{-0.27})$ 		 \\
					&1,000	& 0.1		& & $(0.0109^{+0.0008}_{-0.0007}, -0.19^{+0.04}_{-0.04}, 9.09^{+0.05}_{-0.05}, 1.47^{+0.34}_{-0.31})$ 		  \\
					&1,000	& 0.04	& & $(0.0107^{+0.0006}_{-0.0006}, -0.19^{+0.03}_{-0.03}, 9.08^{+0.04}_{-0.05}, 1.40^{+0.23}_{-0.21})$ 		  \\
\tableline
\end{tabular}
\end{center}
\tablecomments{Columns, from left to right: (1) the accretion model and associated parameters; 
(2) the number of objects in the mock sample; (3) assumed spin measurement error for each 
object in the mock sample; (4) input values of the model parameters to generate the mock sample; 
(5) Bayesian reconstruction results for the model parameters in terms of the mean value and 
$1\sigma$ uncertainties.}
\end{table*}

\section{discussion}
\label{sec:dis}

\subsection{Fitting Mock with a Model Different from the Input One}
\label{sec:dis1}

The results presented in Section~\ref{sec:res} are obtained through MCMC simulations by 
adopting an accretion history model that is the same as the input model to generate the mock 
spin samples. However, the real accretion history is not known a priori and it can only be 
approximated via simplified assumptions on its form (e.g., Model A, B, or C) and constrained 
by matching the model with observations. In the above section, we have demonstrated that the 
accretion model can be well reconstructed if the adopted model is a good approximation to the 
underlying one. Here we also check whether different accretion models can be distinguished 
from each other by fitting the mock sample(s) generated from any one of three models in this 
paper with the other two models.

To compare two competitive models, one may evaluate their relative evidence. The Bayesian 
evidence is the integration of the likelihood over the whole parameter space, i.e., 
\be
p(D|M)=\int p(D|M, {\bm\theta}) p({\bm\theta}|M) d{\bm\theta}, 
\ee
where $M$ denotes the model, $p(D|M, {\bm\theta})$ is the likelihood given by Equation~(\ref{eq-like}), 
and $p({\bm\theta}|M)$ is the prior of the parameters ${\bm\theta}$ associated with the model, which 
is assumed to be a flat distribution. The ratio of Bayesian evidence of two models gives the Bayes factor, 
i.e., $B_{1,2}=p(D|M_1)/p(D|M_2)$. We adopt the sample in the upper panel of Figure~\ref{fig-data3}, 
generated from Model A with $(b, \gamma, f_{\rm c})=(0.012, 0.00, 0.46)$. The sample includes $100$ 
MBHs with a spin accuracy of $0.1$. This is set to check that whether a relatively small sample size and 
low spin accuracy are able to distinguish between different models, since the results shown in Section~\ref{sec:res} indicate that a larger sample with higher spin accuracies will better constrain the model. We then apply 
Model C to fit the data by adopting the same Bayesian MCMC procedure. Although with larger uncertainties, 
the sampling does return peaked posterior distributions with the means and $1\sigma$ errors of $(b, 
\gamma, m_0, \beta)=(0.020^{+0.005}_{-0.004}, -0.14^{+0.09}_{-0.10}, 10.31^{+0.40}_{-0.32}, 
1.21^{+0.89}_{-0.80})$.

However, the Bayes factor we find is $B_{A, C}=19.2$, indicating more evidence for the input Model A 
over Model C.

\subsection{Selection Bias}

Current spin measurement of MBHs relies on the modeling of the reflected X-ray spectra of AGNs, 
and especially the red wing of the Fe K$\alpha$ line, which might suffer from selection biases toward 
high signal-to-noise ratios and thus X-ray-luminous sources \citep[e.g.,][]{Brenneman13}. This may 
lead to a preferred selection of faster-rotating or higher-mass MBHs. In our simulations in 
Section~\ref{sec:res} we do not consider such selection bias, as the number of current spin samples 
is still quite limited and quantitative description of the selection bias is still uncertain. As a simple 
consideration, we select mock spin samples according to Model A with $(b, \gamma, f_{\rm c})=(0.005, 
0.50, 0.20)$, like the lower panel of Figure~\ref{fig-data3} but with a cutoff in bolometric luminosity. With 
a similar sample size, we add a linear dependence of spin accuracy on the MBH mass, i.e., 0.1 for $10^6 
M_\odot$ MBHs and $0.04$ for $10^{10} M_\odot$ MBHs. By adopting the same Bayesian procedure, 
we find that the reconstructed $f_{\rm c}$ is slightly larger than that without considering selection bias 
($0.22\pm0.03$ vs. $0.19\pm0.03$). In the future, once the spin measurements increase by a large 
factor, the selection bias may be estimated with detailed consideration of those systems with/without 
spin measurements. In principle, with such estimations, one may introduce a weight to the spin sample 
to consider the selection bias and obtain more accurate constraints on MBH accretion history.

\subsection{Mergers of MBHs}

Our models do not take MBH mergers into account, since MBHs mainly acquire their masses by means 
of gas accretion, although some works already involved mergers in their models \citep[e.g.,][]{Berti08, 
Barausse12}. Relativistic simulations reveal that mergers of two MBHs with comparable masses generally
result in a spin value of $0.7-0.9$ \citep[e.g.,][]{Berti08, Barausse09}. If a merger occurs in the coherent 
accretion phase, it may lead to a change of the spin from the quasi-equilibrium value, which is mainly 
determined by the disk-to-BH mass ratio. Therefore, it may lead to some bias in the constraint on $b$. 
Because more massive BHs have a higher probability of merging, if a merger event happens during the 
interval of two episodes of chaotic accretion, it will lead to a jump in the MBH spin. After that, the spin 
sharply declines and then oscillates. This may affect the constraint on $f_{\rm c}$. MBH mergers should 
in principle be considered in a more comprehensive cosmological model that involves evolution of both 
MBHs and host galaxies, but beyond the scope of this paper. We defer it to a future work to study the 
effect of mergers on the cosmic spin distributions of MBHs.

\section{conclusions}
\label{sec:con}

In this paper, we investigate constraints on the growth history of MBHs that may be obtained by the spin measurements from X-ray observations in the future, by performing mock simulations and assuming 
simple two-phase accretion models, i.e., MBHs experience coherent accretion followed by periods of 
chaotic accretion. The cloud mass in each chaotic episode is assumed to follow a power-law dependence 
on the MBH mass. The division of the two phases is characterized by parameter $f_{\rm c}$, for which 
we consider three cases: (i) constant for all MBHs; (ii) a power-law dependence on the MBH mass; (iii) 
a hyperbolic-tangent form of dependence on the MBH mass. Those models are capable of producing 
various spin distributions of the mock samples including: (i) high spins for low-mass BHs and low spins 
for high-mass BHs; (ii) low spins for low- and high-mass BHs, and high spins for intermediate-mass BHs; 
(iii) high spins for low- and high-mass BHs, and low spins for intermediate-mass BHs.

By performing Bayesian MCMC simulations and adopting the mock samples generated from various 
accretion models, we find that for the simple model with constant $f_{\rm c}$ for all MBHs, the accretion 
history can be reasonably reproduced provided that spins of $100$ MBHs are measured with an accuracy 
of $0.1$. However, a larger sample (e.g., $300$) is required to remove the tail on the left side of the 
posterior distribution of parameter $f_{\rm c}$. Models with an additional dependence of $f_{\rm c}$ on 
the MBH mass are further explored. With a weak power-law form of $f_{\rm c}$,  the model is not so well reconstructed, especially for parameters associated with $f_{\rm c}$, even a sample of $300$ MBHs is 
provided with a spin accuracy of $0.04$. Instead, the model with a stronger (e.g., hyperbolic-tangent) 
dependence of $f_{\rm c}$ on the MBH mass can be well reconstructed provided a sample has similar
size and spin accuracy.

The effect of a combination of various sample sizes and spin accuracies is additionally explored and 
we find that generally either a larger sample size or a higher spin accuracy will place a more precise 
constraint on the accretion model, and a large sample will somehow compensate for a low spin accuracy 
and vice versa. Future X-ray missions will provide hundreds of spin measurements with a precision of 
$\lesssim0.1-0.2$, and the data are not limited to the local universe \citep[e.g.,][]{Rosa19}, calling for 
improvements in the modeling of MBH growth.

\acknowledgements
We thank the anonymous referee for their helpful suggestions and comments. 
X. Zhang gives the delayed thanks to Massimo Dotti for his kindness and patience 
in replying to the emails and helpful discussions on spin evolution for individual black holes. 
This work is supported by the National Key Program for Science and Technology Research 
and Development No. 2017YFA0402600 and No. 2016YFA0400704, the National Natural 
Science Foundation of China under grants No. 11525312, 11890692, 11873056, 11690024, 
11991052, the Fundamental Research Funds for the Central Universities at Xiamen University 
under grant 20720190115, and the Strategic Priority Research Program of the Chinese Academy 
of Science Multi-wave band Gravitational Wave Universe (grant No. XDB23040000). X. Zhang 
acknowledges the support from the China Postdoctoral Science Foundation (2019M662233).

\end{document}